\documentclass[aps,pra,twocolumn,10pt]{revtex4-1}
\usepackage[utf8]{inputenc}
\usepackage{amsmath,amssymb,graphicx,bm,microtype}
\usepackage[cal=pxtx]{mathalfa}
\usepackage[colorlinks=true,citecolor=blue,urlcolor=blue]{hyperref}

\begin{document}

\title{Van Hove scenario of anisotropic transport in a two-dimensional spin-orbit coupled electron gas in an in-plane magnetic field}

\author{Vladimir A. Sablikov and Yurii Ya. Tkach}

\affiliation{Kotel'nikov Institute of Radio Engineering and Electronics,
Russian Academy of Sciences, Fryazino, Moscow District, 141190, Russia}

\begin{abstract}
We study electronic transport in two-dimensional spin-orbit coupled electron gas subjected to an in-plane magnetic field. The interplay of the spin-orbit interaction and the magnetic field leads to the Van Hove singularity of the density of states and strong anisotropy of Fermi contours. We develop a method that allows one to exactly calculate the nonequilibrium distribution function for these conditions within the framework of the semiclassical Boltzmann equation without using the scattering time approximation. The method is applied to calculate the conductivity tensor and the tensor of spin polarization induced by the electric field (Aronov-Lyanda-Geller-Edelstein effect). It is found that both the conductivity and the spin polarization have a sharp singularity as functions of the Fermi level or magnetic field, which occurs when the Fermi level passes through the Van Hove singularity. In addition, the transport anisotropy dramatically changes near the singularity.
\end{abstract}

\maketitle

\section{Introduction}
The interplay of spin-orbit interaction (SOI) in two-dimensional (2D) electron systems and an in-plane magnetic field attracts significant interest because the magnetic field allows one to manipulate the Fermi contours in a controllable way, which can be an effective tool to study the electronic states and scattering processes. In the general case, in which the magnetic field is directed at an arbitrary angle with respect to the system plane, the spectrum and orbital motion of electrons undergo diverse changes~\cite{winkler}. The in-plane configuration is attractive since the magnetic field does not disturb the orbital wave functions but changes only the Fermi contours due to which it becomes possible to study how the Fermi-contour topology affects the electronic transport. 

Under this condition, there appear such effects as the longitudinal magnetoresistance and the planar Hall effect, which have been widely studied in recent years in many interesting but rather complicated systems. In LaAlO\textsubscript{3}/SrTiO\textsubscript{3} heterojunctions, a strong reconstruction of the Fermi surface occurs near the Lifshitz transition between $d$ orbitals with different symmetry in the presence of an external magnetic field~\cite{joshua2012universal,PhysRevB.87.245121}, due to which a giant magnetoresistance~\cite{PhysRevLett.115.016803} and anisotropic conductivity appear~\cite{PhysRevLett.104.126802,brinkman2007magnetic,PhysRevB.80.140403,Joshua9633,PhysRevB.87.201102,joshua2012universal,PhysRevB.95.205430}. In Al\textsubscript{2}O\textsubscript{3}/SrTiO\textsubscript{3} heterostructures the magnetic field produces a strong additional anisotropy of conductivity~\cite{PhysRevB.95.245132}. The planar Hall effect was recently found in the 2D system of Dirac electrons, which is formed by surface states in topological insulators~\cite{taskin2017planar}. The negative longitudinal magnetoresistance and planar Hall effect are studied also in other situations, where they are not directly related to the SOIs, such as Dirac and Weyl semimetals~\cite{PhysRevB.88.104412,PhysRevLett.113.247203,PhysRevB.91.245157,Xiong413,PhysRevB.96.041110} and even conventional centrosymmetric and time-reversal symmetric semiconductors and metals~\cite{PhysRevLett.120.026601}.

Another aspect of the anisotropic transport in spin-orbit coupled systems is related to anisotropic phases that are formed because of spontaneous breaking of spatial symmetry in strongly interacting electron systems with SOI~\cite{PhysRevB.85.035116,PhysRevB.90.235119,PhysRevB.89.155103,PhysRevB.96.235425,chesiPHD2007}. Though an external magnetic field is absent the anisotropic state of the electron liquid is formed due to a self-consistent magnetic field that is oriented in the plane of the 2D system~\cite{PhysRevB.85.035116}. For these systems the electronic transport anisotropy has not yet been studied.

One of the main problems in the study of the anisotropic transport within the semiclassical approach is that in this case the relaxation-time approximation can not be used to solve the Boltzmann equation~\cite{PhysRevB.79.045427}. In the present paper we develop a method that allows one to solve this problem exactly for arbitrary Fermi contours in the case of elastic scattering of electrons by impurities with short-ranged potential. The method is applied to a model system that makes it possible to study the anisotropy effects for various forms and configurations of Fermi contours. 

We consider a 2D gas of non-interacting electrons with Rashba SOI subjected to an in-plane magnetic field. An important effect produced by the magnetic field is the appearance of a Van Hove singularity of density of states~\cite{Tkach2016} that strongly affects the transport properties and enhances the anisotropy of Fermi contours. We calculate the conductivity tensor and the spin polarization induced by an in-plane electric field (the Aronov-Lyanda-Geller-Edelstein effect~\cite{aronov1989ag,EDELSTEIN1990233}) and study their variation with increasing magnetic field. It is found that i) the conductivity tensor components and the nonequilibrium spin polarization have sharp singularities arising when the Fermi level passes through the Van Hove singularity point, and ii) the axis of the highest conductivity as well as the vector of nonequilibrium spin polarization strongly change their direction near this point.

\section{Electronic states and Fermi contours}\label{S_F_contours}
In this section we present the electronic states and Fermi contours, which will be used to calculate the transport properties. The spectrum of eigenstates of 2D electrons subjected to an in-plane magnetic field in the presence of both Rashba and Dresselhaus SOIs was considered early~\cite{Nakhmedov2012,Tkach2016}. For our purpose, it is enough to restrict ourselves to a simplified situation, in which only the Rashba SOI is present, and to neglect the bands of transverse quantization. Therefore we omit the details and focus on the Fermi contours and density of states.

The Hamiltonian is written as follows:
\begin{equation}
 H=\frac{\mathbf{p}^{2}}{2m}\sigma_0 + \frac{\alpha}{\hbar}(p_x\sigma_y- p_y\sigma_x) - \frac{1}{2}g\mu_B B\sigma_y,
\label{hamiltonian}
\end{equation}
where $\mathbf{p}=(p_x,p_y)$ is the electron momentum, $m$ is the effective mass, $\alpha$ is the SOI constant, $\sigma_{x}$ and $\sigma_{y}$ are the Pauli matrices, $B$ is the magnetic field directed along $y$ axis, $\mu_{B}$ is the Bohr magneton, and $g$ is the effective Land\'e factor, which is supposed to be isotropic and independent of $B$.

There are two types of eigenstates with opposed spins, which we mark with an index $\lambda=\pm$. Their energies and wave functions are
\begin{equation}
 \varepsilon_{\lambda}(\mathbf{k})=k^{2}+2\lambda\sqrt{(k_x-b)^2+k_y^2},
\label{epsilon-k}
\end{equation}
and
\begin{equation}
 \psi_{\mathbf{k}\lambda}(\mathbf{r})= \frac{1}{\sqrt{2A}}\binom{1}{i\lambda e^{i\varphi}} e^{i(k_x x+k_y y)},
\label{wavefunc}
\end{equation}
where here and below we use dimensionless quantities: $\varepsilon$ is the energy normalized to the characteristic energy of the SOI, $E_{so}=m\alpha^2/(2\hbar^2)$; $\mathbf{k}$ is the wave vector normalized to the characteristic wave vector of the SOI, $k_{so}=\alpha m/\hbar^2$; $\mathbf{b}=g\mu_B \mathbf{B}\hbar^2/(2m\alpha^2)$ is the dimensionless magnetic field; $A$ is a normalization area. The phase $\varphi(\mathbf{k})$ is defined by equations 
\begin{equation}
\begin{aligned}
\cos\varphi=&\frac{k \cos\phi-b}{\sqrt{k^2+b^2-2bk\cos\phi}}\,,\\ \sin\varphi=&\frac{k \sin\phi}{\sqrt{k^2+b^2-2bk\cos\phi}}\,,
\end{aligned}
\label{varphi}
\end{equation} 
with $k$ and $\phi$ being the modulus and the azimuthal angle of the wave vector $\mathbf{k}$.

\begin{figure}
\includegraphics[width=.85\linewidth]{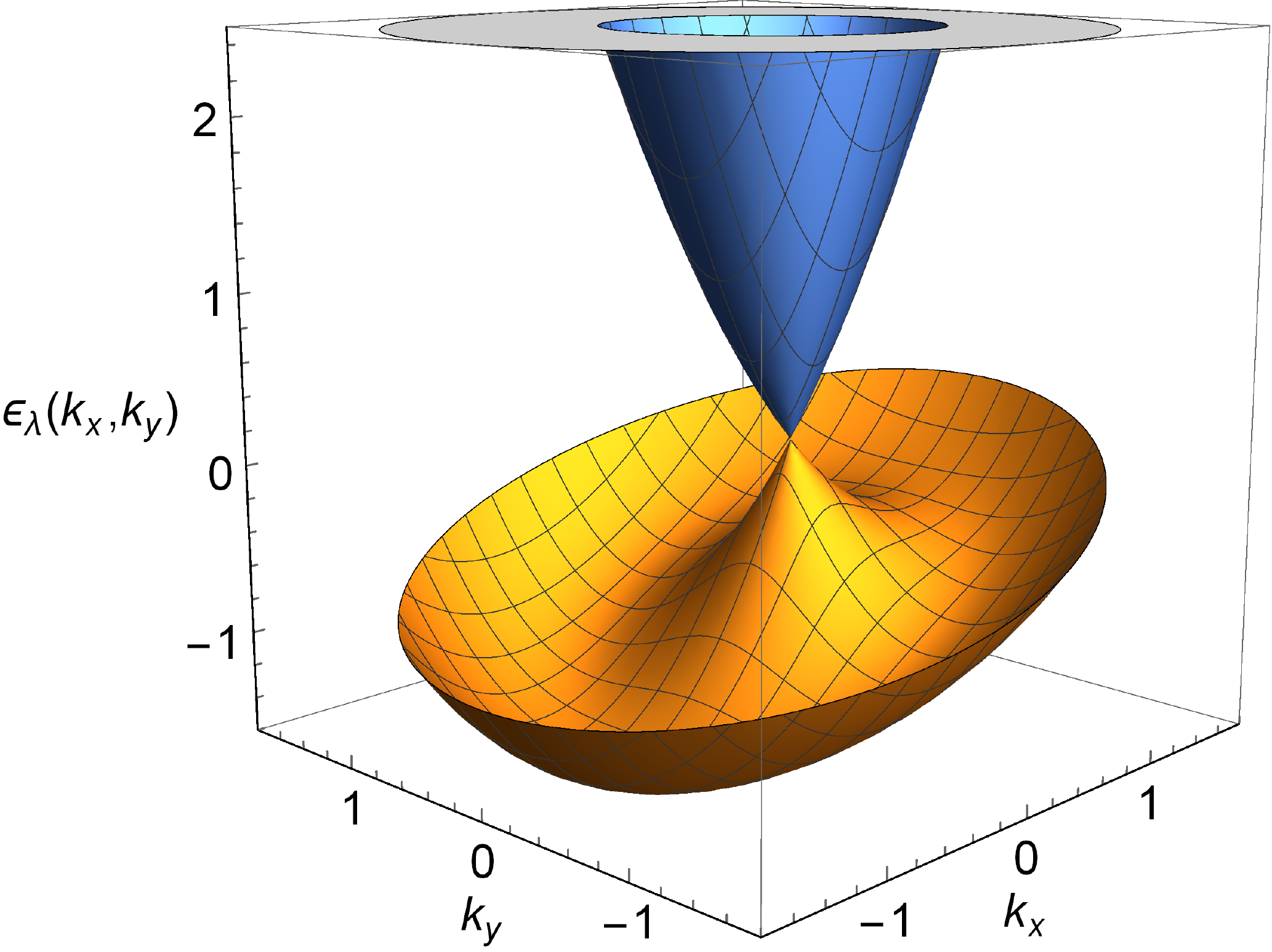}
\caption{(color online) Energy dispersion in the 2D $\mathbf{k}$-space. Blue and orange surfaces are the branches of the spectrum, given by Eq.~(\ref{epsilon-k}), with  $\lambda=+1$ and $\lambda=-1$ for $b=0.3$. The saddle point on the $\lambda=-1$ branch is located at $k_x=1, k_y=0$.}
\label{fig1}
\end{figure}

Due to the presence of a magnetic field the energy dispersion becomes anisotropic in $\mathbf{k}$-space, as is shown in Fig.~\ref{fig1}. The energy landscape in the $\mathbf{k}$-space has a saddle point at $k_x=b/|b|,  k_y=0$. Its energy position is $\varepsilon_{s}=-1+2|b|$. The saddle point exists in the interval of magnetic field $-1<b<1$. The Dirac point is located at the energy $\varepsilon_D=b^2$ in the point ($k_x=b, k_y=0$) of the $\mathbf{k}$-space. With increasing the magnetic field the saddle point comes up from the band bottom to the Dirac point.

\begin{figure}
\includegraphics[width=1.\linewidth]{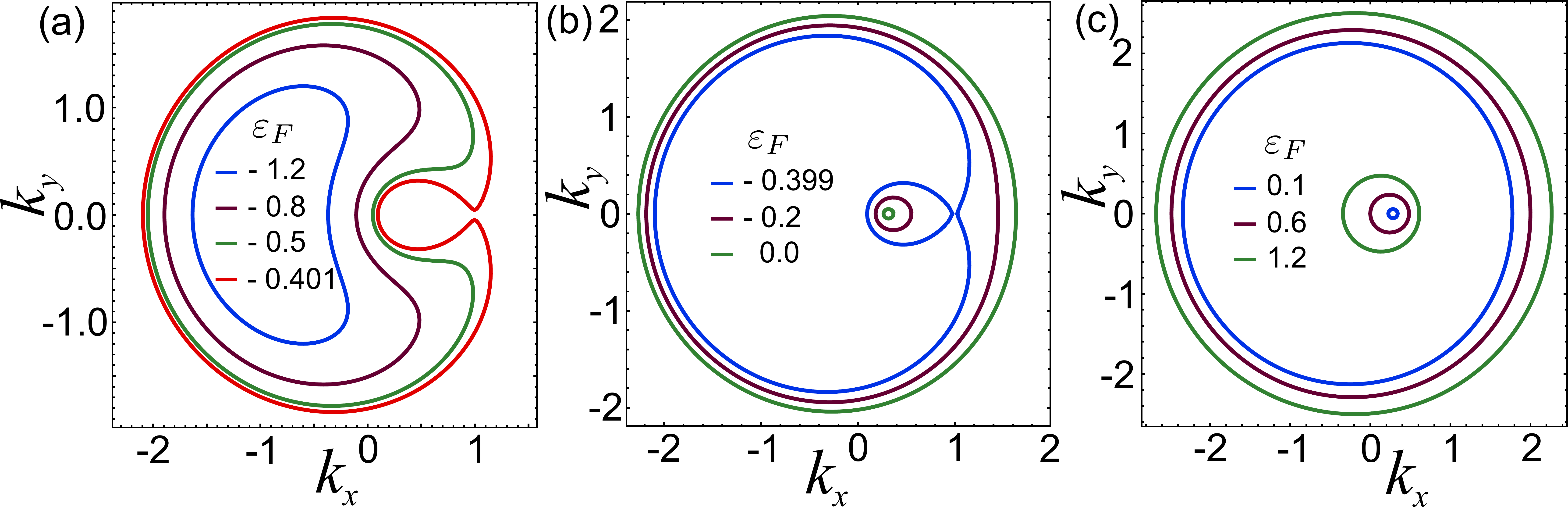}
\caption{(color online) The Fermi contours for a variety of the Fermi energy at magnetic field $b=0.3$ in the following energy regions: (a) below the saddle point, (b) between the saddle point and the Dirac point, and (c) above the Dirac point. The band bottom is located at $\varepsilon_b=-1.6$, the saddle point is $\varepsilon_s=-0.4$, and the Dirac point is $\varepsilon_D= 0.09$.}
\label{fig2}
\end{figure}

In what follows, the Fermi contours will be important. They are defined by the equation $\varepsilon_{\lambda}(\mathbf{k})=\varepsilon_F$. We denote its solutions by $k_{\lambda,r}(\phi)$, where the index $r$ numbers possible solutions at a given $\lambda$. The Fermi contours have a very diverse shape depending on the Fermi energy and magnetic field as shown in Fig.~\ref{fig2}(a-c) for energy regions below the saddle point, above the saddle point but below the Dirac point, and above the Dirac point. 

In the energy range below the Dirac point the Fermi contours are formed by the states with $\lambda=-1$. If the Fermi level lies below the saddle point in the interval, $-1-2|b|<\varepsilon_F<\varepsilon_s$, there is one Fermi contour for a given $\varepsilon_F$, Fig.~\ref{fig2}(a). In the interval $\varepsilon_s<\varepsilon_F<\varepsilon_D$, there are two Fermi contours shown in Fig.~\ref{fig2}(b). When the Fermi energy is above the Dirac point, $\varepsilon_D<\varepsilon_F$, each branch of the spectrum with $\lambda=\pm 1$ gives one contour demonstrated in Fig.~\ref{fig2}(c). 

In the saddle point, the density of states has logarithmic Van Hove singularity, $D(\varepsilon)\sim \ln|\varepsilon-\varepsilon_s|$, in accordance with the general theory~\cite{PhysRev.89.1189}. As the magnetic field goes to zero, the saddle point disappears and the van Hove singularity transforms to the well known singularity $D(\varepsilon)\sim(\varepsilon+1)^{-1/2}$ at the band bottom. In the interval $0<|b|<1$, the energy of the Van Hove singularity is controlled by the magnetic field.

It is therefore interesting to trace how the transport properties change when the Fermi level is changed at a given magnetic field, and also when the magnetic field is changed at a given Fermi level. The first regime, in which the magnetic field is fixed, has the advantage that the energy dispersion $\varepsilon_{\lambda}(\mathbf{k})$ remains unchanged when scanning the Fermi level. It can be expected that the transport properties have a sharp feature when the Fermi level crosses the Van Hove singularity. The transport anisotropy can also change dramatically since the saddle point is located asymmetrically with respect to the center of the Brillouin zone. 

\section{Boltzmann equation}\label{S_kin_equation}
Now we turn to the transport induced by an in-plane electric field. The electron current and spin polarization are studied using the semiclassical Boltzmann equation. For a small homogeneous electric field $\boldsymbol{\mathcal{E}}$, the distribution function $f(\mathbf{k})$ is determined by the Boltzmann equation, which we present in a familiar linear form~\cite{0034-4885-62-2-004} 
\begin{multline}
 -e \boldsymbol{\mathcal{E}}\mathbf{v}_{\lambda}(\mathbf{k})\bigl(-\partial_{\varepsilon}f_0(\varepsilon)\bigr) \\
 = \sum_{\lambda'}\int \frac{d^2k'}{4\pi^2} W_{\lambda \mathbf{k},\lambda' \mathbf{k}'} \bigl[f_{\lambda}(\mathbf{k},\boldsymbol{\mathcal{E}})-f_{\lambda'}(\mathbf{k}',\boldsymbol{\mathcal{E}})\bigr],
\label{Boltzmann}
\end{multline}
where $\mathbf{v}_{\lambda}(\mathbf{k})$ is the group velocity in the state $|\lambda,\mathbf{k}\rangle$, $f_0$ is the equilibrium distribution function, and $W_{\lambda \mathbf{k},\lambda' \mathbf{k}'}$ is the scattering probability.

The anisotropy of the electron dispersion leads to the scattering anisotropy, because of which the collision integral on the right-hand side of Eq.~\ref{Boltzmann} can not be simplified by introducing a relaxation time. This problem was debated in the recent literature~\cite{Schwab2002,PhysRevB.68.165311,PhysRevB.75.155323,PhysRevB.77.115352,PhysRevB.79.045427}. Using a relaxation time that depends on the direction of $\mathbf{k}$ is certainly not precise though it captures some aspects of the anisotropic transport. An improved approach based on introducing two relaxation times, $\tau_{\|}$ and $\tau_{\bot}$, was proposed by Schliemann and Loss~\cite{PhysRevB.68.165311} and applied to the study of anisotropic transport and magnetotransport in a 2D electron gas in the presence of both Rashba and Dresselhaus SOIs~\cite{PhysRevB.68.165311,PhysRevB.75.155323}. However, V\'yborn\'y et al~\cite{PhysRevB.79.045427} have shown that this approach not only fails to exactly calculate the distribution function, but in some cases leads to erroneous results. Instead, the Boltzmann equation should be solved as an integral equation. We proceed in this way.

We consider the scattering by impurities with short-ranged potential approximated as $V(r)=V_0\delta(\mathbf{r})$. The impurity concentration $N$ is assumed to be small, so that their potentials do not overlap and the scattering processes from different impurities are not correlated. 

Having calculated the scattering probability in the Born approximation by using the wave functions (\ref{wavefunc}) we arrive at the following equation for the nonequilibrium part of the distribution functions $\Delta f_{\lambda}(\mathbf{k})$:
\begin{multline}
\sum\limits_{\lambda'}\int\frac{d^2k'}{\pi}\bigl(1+\lambda\lambda'\cos[\varphi(\mathbf{k})-\varphi(\mathbf{k}')]\bigr)\delta\bigl(\varepsilon_{\lambda}(\mathbf{k})-\varepsilon_{\lambda'}(\mathbf{k}')\bigr) \\\times \bigl[\Delta f_{\lambda}(\mathbf{k})-\Delta f_{\lambda'}(\mathbf{k}')\bigr]=\frac{e\boldsymbol{\mathcal{E}}\mathbf{v}_{\lambda}(\mathbf{k})}{R}\frac{\partial f_0}{\partial \varepsilon} \,,
\label{Boltzmann1}
\end{multline}
where the dimensionless variables are used. The electric field $\boldsymbol{\mathcal{E}}$ is normalized to $E_{so}k_{so}/e$, the group velocity is $\mathbf{v}_{\lambda}=\nabla_{\mathbf{k}}\varepsilon_{\lambda}$, and $R$ is a unique parameter that appears for this system: $R=V_0^2N/\alpha^2$.

The nonequlibrium distribution function $\Delta f_{\lambda}(\mathbf{k})$ can be written in the form
\begin{equation}
\Delta f_{\lambda}(\mathbf{k})= \frac{e\mathcal{E}}{R} \mathcal{F}_{\lambda}(\mathbf{k})\frac{\partial f_0}{\partial \varepsilon}\,.
\label{delta_f}
\end{equation}
The function $\mathcal{F}_{\lambda}(\mathbf{k})$ introduced here is determined by an equation that can be readily obtained in the case of zero temperature by integrating with respect to the modulus of $\mathbf{k}$ in Eq.~(\ref{Boltzmann1}). In this case the integration is carried out along the Fermi contours. Since in some cases the contours have a complicated shape, such that $k(\phi)$ is in general a multivalued function of $\phi$, we have to divide the contour into parts within which $k(\phi)$ is a single-valued function. Each part will be marked by index $r$ which can vary from 1 to 4 depending on the contour shape. Keeping this in mind we will add this index to the notations of the functions and integrals defined on the Fermi contours. 

The function $\mathcal{F}_{\lambda,r}(\mathbf{k})$ is defined on the corresponding part of the Fermi contour $k=k_{\lambda,r}(\phi)$ by the following equations:
\begin{multline}
\sum\limits_{\lambda',r'}\int\frac{d\phi'}{\pi}\bigl(1+\lambda\lambda'\cos[\varphi_{\lambda,r}(\phi)-\varphi_{\lambda',r'}(\phi')]\bigr)M_{\lambda',r'}(\phi')\\ \times \bigl[\mathcal{F}_{\lambda,r}(\phi)-\mathcal{F}_{\lambda',r'}(\phi')\bigr]=\mathcal{G}_{\lambda,r}(\phi)\,,
\label{eq_Flr}
\end{multline}
where 
\begin{equation}
M_{\lambda,r}(\phi)=\Bigl[k\Bigm/\frac{\partial \varepsilon_{\lambda}(\mathbf{k})}{\partial k}\Bigr]_{k=k_{\lambda,r}(\phi)},
\label{M}
\end{equation}
\begin{equation}
\varphi_{\lambda,r}(\phi)=\varphi(\mathbf{k})\Bigr|_{k=k_{\lambda,r}(\phi)}\,,
\label{Phi}
\end{equation}
\begin{equation}
\mathcal{G}_{\lambda,r}(\phi)=\frac{v_{\lambda}(\mathbf{k})\cos[\xi(\mathbf{k})-\theta]}{R}\Biggr|_{k=k_{\lambda,r}(\phi)}\,.
\label{G}
\end{equation}
Here $\xi(\mathbf{k})$ is the angle between $\mathbf{v}_{\lambda,r}(\phi)$ and the $x$-axis, and $\theta$ is the angle between $\boldsymbol{\mathcal{E}}$ and the $x$-axis. The values $M_{\lambda,r}(\phi)$, $\varphi_{\lambda,r}(\phi)$ and $\mathcal{G}_{\lambda,r}(\phi)$ are defined on the corresponding Fermi contours. They are easily calculated using Eqs.~(\ref{epsilon-k}) and (\ref{wavefunc}).

Equation~(\ref{eq_Flr}) is solved analytically since it is merely a linear Fredholm equation with a degenerate kernel. We represent the kernel as the sum of the products of two functions, one of which is a function of $\phi$, and the other is a function of $\phi'$. In our case, these functions are simply sines and cosines. As a result we arrive at the following form of the function $\mathcal{F}_{\lambda,r}$:
\begin{equation}
\mathcal{F}_{\lambda,r}(\phi,\theta)\!=\!\frac{\mathcal{G}_{\lambda,r}(\phi)\!+\!\mathfrak{A}\!+\!\lambda\mathfrak{B}\cos[\varphi_{\lambda,r}(\phi)]\!+\!\lambda\mathfrak{C}\sin[\varphi_{\lambda,r}(\phi)]}{A+\lambda B\cos[\varphi_{\lambda,r}(\phi)]+\lambda C\sin[\varphi_{\lambda,r}(\phi)]},
\label{Flr}
\end{equation}
where the coefficients $A$, $B$, $C$, $\mathfrak{A}$, $\mathfrak{B}$ and $\mathfrak{C}$ are directly determined from Eq.~(\ref{eq_Flr}),
\begin{align}
A &=\sum\limits_{\lambda,r}\int\frac{d\phi}{\pi} M_{\lambda,r}(\phi)\,,\\
B &=\sum\limits_{\lambda,r}\lambda\int\frac{d\phi}{\pi} M_{\lambda,r}(\phi)\cos[\varphi_{\lambda,r}(\phi)]\,,\\
C &=\sum\limits_{\lambda,r}\lambda\int\frac{d\phi}{\pi} M_{\lambda,r}(\phi)\sin[\varphi_{\lambda,r}(\phi)]\,;
\label{ABC}
\end{align}
\begin{align}
\mathfrak{A}(\theta) &=\sum\limits_{\lambda,r}\int\frac{d\phi}{\pi} M_{\lambda,r}(\phi)\mathcal{F}_{\lambda,r}(\phi,\theta)\,,\label{A-F}\\
\mathfrak{B}(\theta) &=\sum\limits_{\lambda,r}\lambda\int\frac{d\phi}{\pi} M_{\lambda,r}(\phi)\mathcal{F}_{\lambda,r}(\phi,\theta)\cos[\varphi_{\lambda,r}(\phi)]\,,\label{B-F}\\
\mathfrak{C}(\theta) &=\sum\limits_{\lambda,r}\lambda\int\frac{d\phi}{\pi} M_{\lambda,r}(\phi)\mathcal{F}_{\lambda,r}(\phi,\theta)\sin[\varphi_{\lambda,r}(\phi)]\,.
\label{C-F}
\end{align}

The coefficients $A$, $B$, and $C$ can be straightforwardly calculated if the electron dispersion is known. However, the coefficients $\mathfrak{A}$, $\mathfrak{B}$ and $\mathfrak{C}$ are defined by integrals containing the function $\mathcal{F}_{\lambda,r}(\phi,\theta)$ which is still unknown. To obtain equations that allow one to find the coefficients $\mathfrak{A}$, $\mathfrak{B}$ and $\mathfrak{C}$, we substitute Eq.~(\ref{Flr}) into Eqs.~(\ref{A-F}), (\ref{B-F}), and (\ref{C-F}). As a result, we arrive at a system of linear algebraic equations for these coefficients. In this way the problem is solved.

Further calculations are simplified using the symmetry properties of the Fermi contours and the integrands. In the specific situation considered in this paper, the values $k_{\lambda,r}(\phi)$, $M_{\lambda,r}(\phi)$ and $\cos\varphi_{\lambda,r}(\phi)$ are even functions of $\phi$. The values $\sin\varphi_{\lambda,r}(\phi)$ and $\sin\xi_{\lambda,r}(\phi)$ are odd functions. In this case, $C=0$ and the function $\mathcal{F}_{\lambda,r}(\phi,\theta)$ is simplified to
\begin{multline}
\mathcal{F}_{\lambda,r}(\phi,\theta)=\Bigl\{v_{\lambda,r}\cos\left[\theta-\xi_{\lambda,r}(\phi)\right]+\mathfrak{A}+\lambda\mathfrak{B}\cos[\varphi_{\lambda,r}(\phi)]\\+\lambda\mathfrak{C}\sin[\varphi_{\lambda,r}(\phi)]\Bigr\}D_{\lambda,r}^{-1}(\phi),
\label{Flr1}
\end{multline}
where
\begin{equation}
D_{\lambda,r}(\phi)=A+\lambda B\cos[\varphi_{\lambda,r}(\phi)]\,.
\end{equation}

The system of equations for the coefficients $\mathfrak{A}$, $\mathfrak{B}$ and $\mathfrak{C}$ splits into a separate system of two equations for $\mathfrak{A}$ and $\mathfrak{B}$ and a separate equation for $\mathfrak{C}$. 

The coefficients $\mathfrak{A}$ and $\mathfrak{B}$ are defined by the equations:
\begin{align}
(1-a_0)\mathfrak{A}(\theta)-a_1\mathfrak{B}(\theta)& =g_0\cos\theta \\
-a_1\mathfrak{A}(\theta)+(1-a_2)\mathfrak{B}(\theta)& =g_1\cos\theta\,.
\label{A-B_system}
\end{align}

The coefficient $\mathfrak{C}$ equals to
\begin{equation}
\mathfrak{C}(\theta)=\mathfrak{C}_0\sin\theta \,,
\label{C_1}
\end{equation}
where
\begin{equation}
\mathfrak{C}_0=\frac{g_2}{1-b_2} \,.
\label{C_0}
\end{equation}
Here the following notations are introduced:
\begin{equation}
\begin{aligned}
a_0 = &\sum\limits_{\lambda,r}\int\frac{d\phi}{\pi}\frac{M_{\lambda,r}(\phi)}{D_{\lambda,r}(\phi)},\\
a_1 = &\sum\limits_{\lambda,r}\lambda\int\frac{d\phi}{\pi}\frac{M_{\lambda,r}(\phi)}{D_{\lambda,r}(\phi)}\cos[\varphi_{\lambda,r}(\phi)],\\
a_2 = &\sum\limits_{\lambda,r}\int\frac{d\phi}{\pi}\frac{M_{\lambda,r}(\phi)}{D_{\lambda,r}(\phi)}\cos^2[\varphi_{\lambda,r}(\phi)],\\
b_2 = &\sum\limits_{\lambda,r}\int\frac{d\phi}{\pi}\frac{M_{\lambda,r}(\phi)}{D_{\lambda,r}(\phi)}\sin^2[\varphi_{\lambda,r}(\phi)],
\end{aligned}
\label{a_i}
\end{equation}
\begin{equation}
\begin{aligned}
g_0 = &\sum\limits_{\lambda,r}\int\frac{d\phi}{\pi}\frac{M_{\lambda,r}(\phi)v_{\lambda,r}(\phi)}{D_{\lambda,r}(\phi)}\cos[\xi_{\lambda,r}(\phi)],\\
g_1 = &\sum\limits_{\lambda,r}\lambda\int\frac{d\phi}{\pi}\frac{M_{\lambda,r}(\phi)v_{\lambda,r}(\phi)}{D_{\lambda,r}(\phi)}\cos[\varphi_{\lambda,r}(\phi)]\cos[\xi_{\lambda,r}(\phi)],\\
g_2 =&\sum\limits_{\lambda,r}\lambda\int\frac{d\phi}{\pi}\frac{M_{\lambda,r}(\phi)v_{\lambda,r}(\phi)}{D_{\lambda,r}(\phi)}\sin[\varphi_{\lambda,r}(\phi)]\sin[\xi_{\lambda,r}(\phi)].
\end{aligned}
\label{g_i}
\end{equation}

It is easy to show that there are simple relations between the coefficients $a_0$, $a_1$ and $a_2$, and the coefficients $g_0$ and $g_1$ that simplify the calculations:
\begin{equation}
a_1=\frac{A}{B}(1-a_0),\quad a_2=1-\frac{A^2}{B^2}(1-a_0),
\label{a_i_relations}
\end{equation}
\begin{equation}
g_1=-\frac{A}{B}g_0.
\label{g_i_relations}
\end{equation}
One should note that these relations are proven by analyzing only Eq.~(\ref{a_i}), (\ref{g_i}) and (\ref{ABC}) without using any specific dispersion equation. Therefore, they are quite general.

A peculiarity of the equation system~(\ref{A-B_system}) is that both equations are not independent. This follows directly from the above relations~(\ref{a_i_relations}) and (\ref{g_i_relations}). It is seen that the equations differ only by a factor. This means that Eqs.~(\ref{A-B_system}) establish only a relation between $\mathfrak{A}$ and $\mathfrak{B}$ and an additional equation is required to determine both coefficients. It is clear that this equation should be obtained from the requirement of electroneutrality of the system, which is not violated under nonequilibrium conditions considered here. The fact that this additional condition is necessary is not surprising, since the Boltzmann equation in the form~(\ref{Boltzmann}) does not automatically guarantee the neutrality. So we should use the equation
\begin{equation}
\sum\limits_{\lambda,r}\int d\phi M_{\lambda,r}(\phi) \mathcal{F}_{\lambda,r}(\phi,\theta)=0\,,
\label{neutrality}
\end{equation}
which allows one to determine the coefficients $\mathfrak{A}$ and $\mathfrak{B}$.

Thus, using Eqs.~(\ref{A-B_system}) and (\ref{neutrality}) we find
\begin{equation}
\mathfrak{A}=0\,,\quad \mathfrak{B}=\mathfrak{B}_0\cos\theta\, 
\end{equation}
where
\begin{equation}
\mathfrak{B}_0=-\frac{g_0}{a_1}\,.
\label{B_0}
\end{equation}

Thus the distribution function in the final form reads
\begin{equation}
\begin{split}
\mathcal{F}_{\lambda,r}(\phi,\theta)&=\frac{v_{\lambda,r}(\phi)\sin[\xi_{\lambda,r}(\phi)]+\lambda\mathfrak{C}_0\sin[\varphi_{\lambda,r}(\phi)]}{D_{\lambda,r}(\phi)}\sin\theta \\
&+\frac{v_{\lambda,r}(\phi)\cos[\xi_{\lambda,r}(\phi)]+\lambda\mathfrak{B}_0\cos[\varphi_{\lambda,r}(\phi)]}{D_{\lambda,r}(\phi)}\cos\theta.
\end{split}
\label{Flr-fin}
\end{equation}
Here the first term on the right hand side is an odd function of $\phi$ and the second term is an even one. The coefficients $\mathfrak{B}_0$ and $\mathfrak{C}_0$ are defined by Eqs.~(\ref{B_0}) and (\ref{C_0}).

Note that in contrast to the method developed in Ref.~\cite{PhysRevB.79.045427} we find the angular dependence of the distribution function exactly. There is no need to expand it in the Fourier series and calculate harmonics. Thus, we can take fully into account the asymmetry of the scattering processes for arbitrary Fermi contours and include electron transitions both within each contour and between different contours in the $\mathbf{k}$-space. 

To test this approach, we have calculated the conductivity and the spin polarization in the case of zero magnetic field when the system is isotropic, but the SOI essentially affects electron scattering processes. This problem was considered in the recent literature~\cite{PhysRevLett.116.166602,PhysRevB.93.195440}. In particular, it was found that the conductivity deviated from the standard Drude’s law if the Fermi level lay below the Dirac point~\cite{PhysRevLett.116.166602}. Using our approach greatly simplifies the calculations and leads to the results coinciding with those of Refs.~\onlinecite{PhysRevLett.116.166602,PhysRevB.93.195440}.

\section{Anisotropic transport}\label{S_responses}
\subsection{Conductivity tensor}\label{S_conductivity}
Conductivity is calculated in the standard way. Using Eq.~(\ref{Flr-fin}) for the distribution function and taking into account the symmetry relations for $k_{\lambda,r}$, $M_{\lambda,r}$, $\varphi_{\lambda,r}$, and $\xi_{\lambda,r}$ with respect to $\phi$, we get the following tensor of the conductivity in dimensionless form:
\begin{align}
G_{xx}=&\sum\limits_{\lambda,r}\int\frac{d\phi}{2\pi}\frac{M_{\lambda,r}(\phi)v_{\lambda,r}(\phi)}{D_{\lambda,r}(\phi)}\cos[\xi_{\lambda,r}(\phi)]\notag\\ \times&\left\{v_{\lambda,r}(\phi)\cos[\xi_{\lambda,r}(\phi)]+\lambda\mathfrak{B}_0\cos[\varphi_{\lambda,r}(\phi) \right\},\label{Gxx}\\
G_{yy}=&\sum\limits_{\lambda,r}\int\frac{d\phi}{2\pi}\frac{M_{\lambda,r}(\phi)v_{\lambda,r}(\phi)}{D_{\lambda,r}(\phi)}\sin[\xi_{\lambda,r}(\phi)]\notag\\ \times&\left\{v_{\lambda,r}(\phi)\sin[\xi_{\lambda,r}(\phi)]+\lambda\mathfrak{C}_0\sin[\varphi_{\lambda,r}(\phi) \right\},
\label{Gyy}
\end{align}
with the conductivity being normalized to $e^2/(hR)$. The off-diagonal components are absent, $G_{xy}=G_{yx}=0$. 

If the electric field is directed at the angle $\theta\ne n\pi/2$ a planar Hall effect appears, with the Hall resistance being a $\pi$-periodic function of the angle between the electric and magnetic fields.

It is not difficult to analyze analytically Eqs.~(\ref{Gxx}) and (\ref{Gyy}) in the limiting case when the Fermi level is near the bottom of the band, $\varepsilon_F=-(1+2|b|)+\Delta\varepsilon$, by expanding in $\Delta\varepsilon$. We get the following result:
\begin{equation}
G_{xx}\approx\frac{\Delta\varepsilon}{2}\,,\quad G_{yy}\approx\frac{\Delta\varepsilon}{2}\frac{b}{1+b}\,.
\label{G_approx}
\end{equation}
Hereinafter, we assume for definiteness that $b>0$. As can be seen, the ratio of conductivities $G_{xx}$ and $G_{yy}$ is $G_{xx}/G_{yy}=b/(1+b)$. This result is easy to understand, given that according to Eq.~(\ref{epsilon-k}) the electron dispersion near the band bottom is
\begin{equation}
\Delta\varepsilon\approx q_x^2+\frac{b}{1+b}q_y^2\,,
\end{equation}
with $q_x$ and $q_y$ being the wave vector components measured with respect to the energy minimum. Thus, the  ratio of conductivities is exactly equal to the ratio of the components of the effective-mass tensor, as one would expect.

The conductivity tensor components $G_{xx}$ and $G_{yy}$ are studied in more details using direct numerical calculations of the integrals in Eqs.~(\ref{Gxx}), (\ref{Gyy}). Below, we present the results of our calculations of conductivity for two regimes: when the Fermi energy is changed at a fixed magnetic field and when the magnetic field is changed while the Fermi level remains unchanged. 

\begin{figure}
\includegraphics[width=1.0\linewidth]{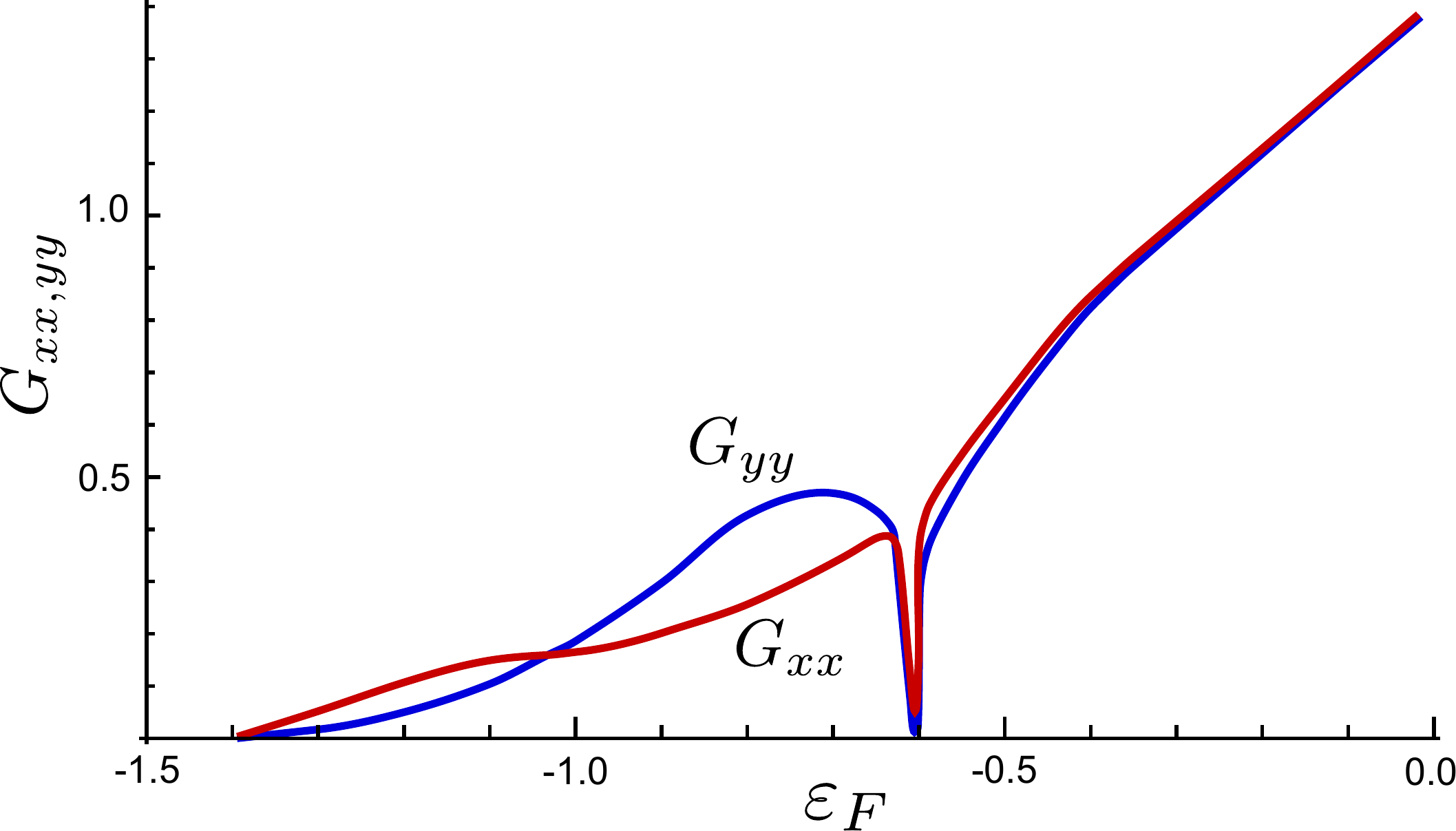}
\caption{(color online) The components of the conductivity tensor as functions of the Fermi energy at $b=0.2$.}
\label{fig3}
\end{figure}

First consider $G_{xx}$ and $G_{yy}$ as functions of $\varepsilon_F$. The results obtained for magnetic field $b=0.2$ are shown in Fig.~\ref{fig3}. In this case, the lower boundary of the band spectrum is $\varepsilon_b=-1.4$, the saddle point is located at $\varepsilon_s=-0.6$, and the Dirac point is $\varepsilon_D=0.04$. It is seen that the most striking effect is a sharp dip in conductivity, which occurs when the Fermi level is near the saddle point. Another interesting feature is that the conductivity is strongly anisotropic. The anisotropy can be characterized by a value $\delta=(G_{xx}-G_{yy})/(G_{xx}+G_{yy})$. Near the band bottom, $\delta\approx 0.7$. As $\varepsilon_F$ increases, the anisotropy changes sign, and when $\varepsilon_F$ approaches the saddle point, $\delta$ reaches -0.3. In the immediate vicinity of the saddle point, $\delta$ changes sign again, and then decreases monotonically.

The sharp drop in the conductivity near the saddle point is undoubtedly caused by the strong increase in the scattering rate because of the Van Hove singularity. The strong anisotropy of the conductivity is caused by anisotropic distribution of the density of states and electron velocity in  $\mathbf{k}$-space. The anisotropy is particularly large when the electron energy is not high in comparison with the Zeeman energy and the characteristic energy of the SOI. An important role in the formation of transport anisotropy is played by the Van Hove singularity. Since the saddle point is shifted along the $x$ axis in $\mathbf{k}$-space relative to the band center, the Fermi contours become strongly anisotropic when the energy is close to the saddle point as can be seen from Fig.~\ref{fig2}. This leads to a strong anisotropy of electron scattering, due to which the conductivity along the magnetic field becomes predominant as $\varepsilon_F$ approaches the saddle point, but is below it.

A noticeable feature is that the presence of a saddle point in the energy dispersion stimulates the appearance of the conductivity anisotropy of a different sign compared to the anisotropy that exists outside the saddle region. As a result, the anisotropy changes sign twice: below the saddle point ($\varepsilon_F=-1.03$ in Fig.~\ref{fig3}) and in close proximity to the point singularity.

In the energy region above the Dirac point, the anisotropy decreases rapidly and its magnitude is very small. This behavior of the anisotropy is consistent with that found in Ref.~\cite{Schwab2002}. At $\varepsilon_F\sim 1$, the anisotropy is estimated on the level $\delta \sim 10^{-3}$ and therefore seems not to be interesting.  

\begin{figure}
\includegraphics[width=1.0\linewidth]{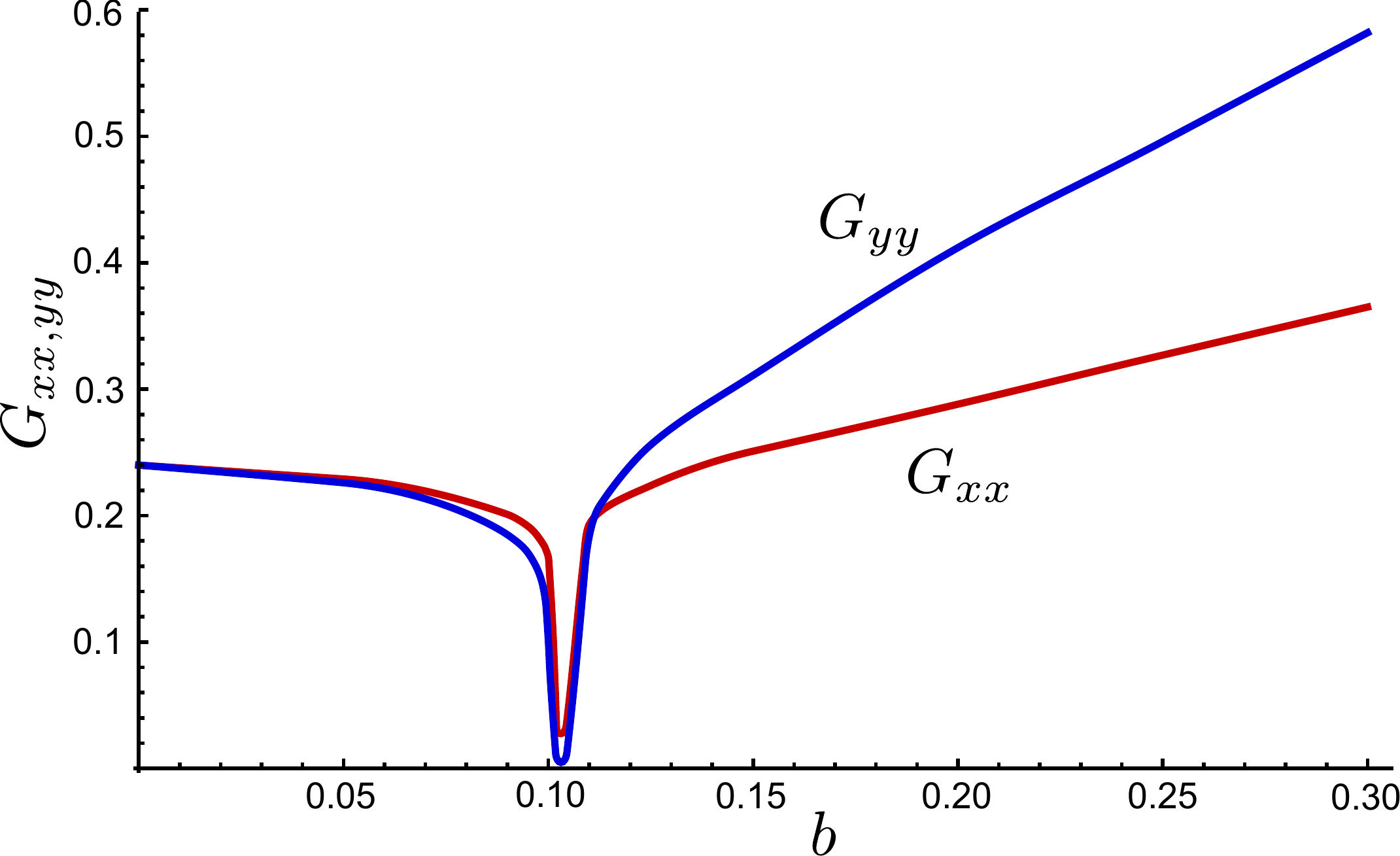}
\caption{(color online) The conductivity tensor components as functions of the magnetic field for the Fermi energy below the Dirac point $\varepsilon_F=-0.8$.}
\label{fig4}
\end{figure}

The dependence of conductivity on the magnetic field at a fixed Fermi level has qualitatively similar features. This is easy to understand keeping in mind that the position of the saddle  point changes upon changing the magnetic field, and it can intersect the Fermi level. The conductivity tensor components calculated as functions of $b$ are shown in Fig.~\ref{fig4}.

\subsection{Spin polarization}\label{S_spin}
The spin density induced by an electric field is calculated using the nonequilibrium distribution function as follows
\begin{equation}\label{spin}
S_{i}=\frac{\hbar}{2}\sum_{\lambda}\int \frac{d^2k}{4\pi^2}\langle\psi^{\dag}_{\lambda,\mathbf{k}}|\sigma_{i}|\psi_{\lambda,\mathbf{k}}\rangle\Delta f_{\lambda}(\mathbf{k})\,,
\end{equation}
where $S_{i}$ is the spin density component, $i=(x,y,z)$, and $\sigma_{i}$ are the Pauli matrices. The spin susceptibility with respect to electric field $\chi_{i j}$, which is often called the Edelstein conductivity, is defined by the equation:
\begin{equation}
S_{i}=\sum_{j}\chi_{i j}\mathcal{E}_{j}\,.
\end{equation}

Using Eqs.~(\ref{wavefunc}), (\ref{delta_f}), (\ref{Flr-fin}) and the symmetry properties of the integrands, we arrive at the following expressions for the Edelstein conductivity in dimensionless form:
\begin{align}
\chi_{xy}=&-\sum\limits_{\lambda,r}\lambda\int\frac{d\phi}{2\pi}\frac{M_{\lambda,r}(\phi)}{D_{\lambda,r}(\phi)}\sin[\varphi_{\lambda,r}(\phi)]\notag\\ \times&\left\{v_{\lambda,r}(\phi)\sin[\xi_{\lambda,r}(\phi)]+\lambda\mathfrak{C}_0\sin[\varphi_{\lambda,r}(\phi) \right\},\label{chi_xy}\\
\chi_{yx}=&\sum\limits_{\lambda,r}\lambda\int\frac{d\phi}{2\pi}\frac{M_{\lambda,r}(\phi)}{D_{\lambda,r}(\phi)}\cos[\varphi_{\lambda,r}(\phi)]\notag\\ \times&\left\{v_{\lambda,r}(\phi)\cos[\xi_{\lambda,r}(\phi)]+\mathfrak{A}_0+\lambda\mathfrak{B}_0\cos[\varphi_{\lambda,r}(\phi) \right\}.
\label{chi_yx}
\end{align}
The diagonal components are zero, $\chi_{xx}=\chi_{yy}=0$. Here the Edelstein conductivity is normalized to $e\hbar/(2\pi\alpha R)$. The $z$-component of the spin polarization is absent, $S_z=0$.

The main features of the Edelstein conductivity are similar to those of the conductivity $G$, but the anisotropy is stronger. 

When the Fermi level is near the band bottom, the components of the Edelstein-conductivity tensor are approximated as follows: 
\begin{equation}
\chi_{xy}\approx \frac{\Delta \varepsilon}{4(1+b)},\quad \chi_{yx}=O(\Delta \varepsilon^2).
\end{equation}
The component $\chi_{xy}$ is seen to be much larger than $\chi_{yx}$. This fact is explained by the strong equilibrium spin polarization of electrons in the $y$-direction, which is caused by the external magnetic field.

\begin{figure}
\includegraphics[width=1.0\linewidth]{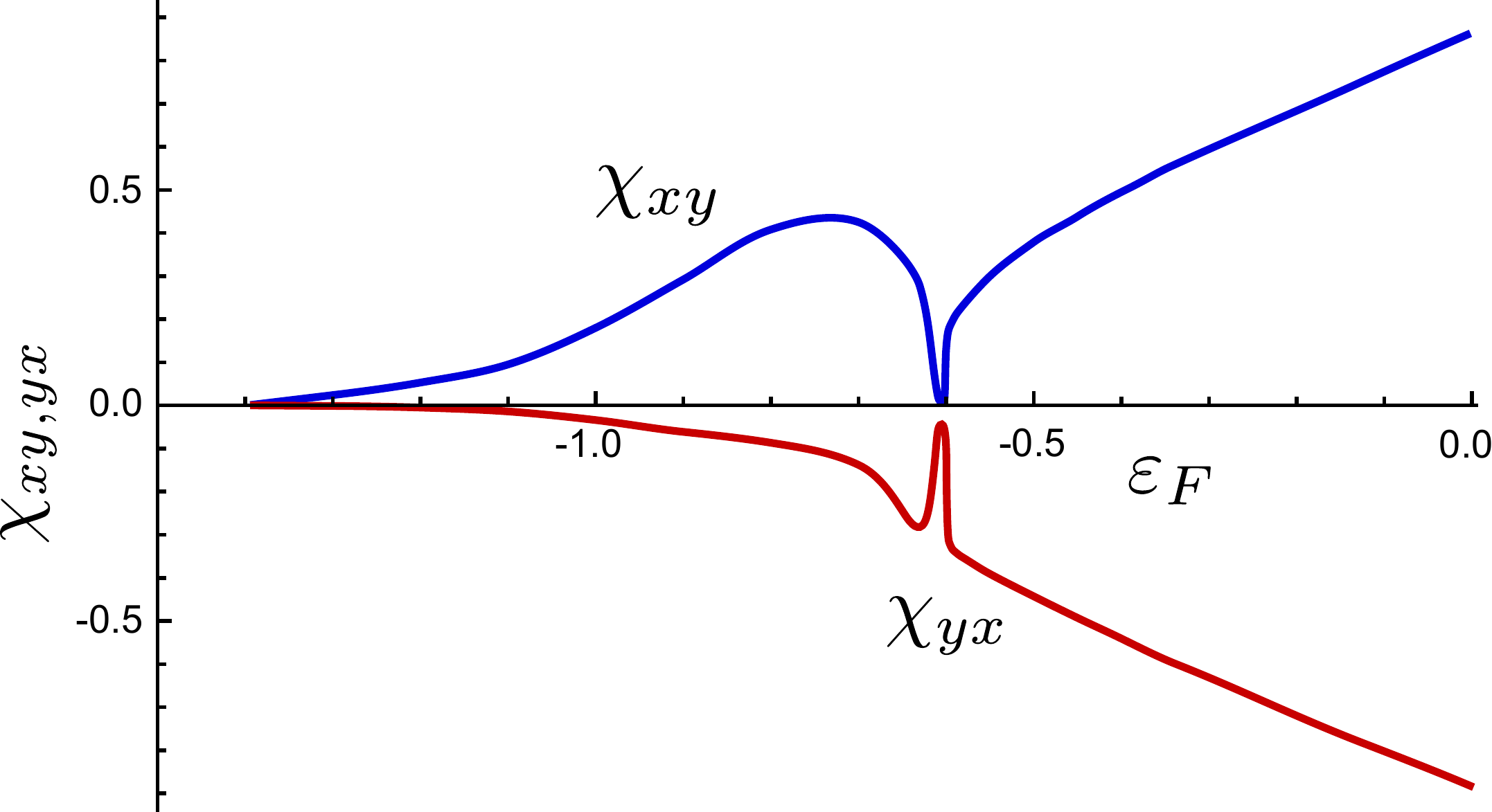}
\caption{(color online) The off-diagonal components of the Edelstein conductivity as functions of the Fermi energy below the Dirac point at fixed magnetic field $b=0.2$.}
\label{fig5}
\end{figure}

\begin{figure}
\includegraphics[width=1.0\linewidth]{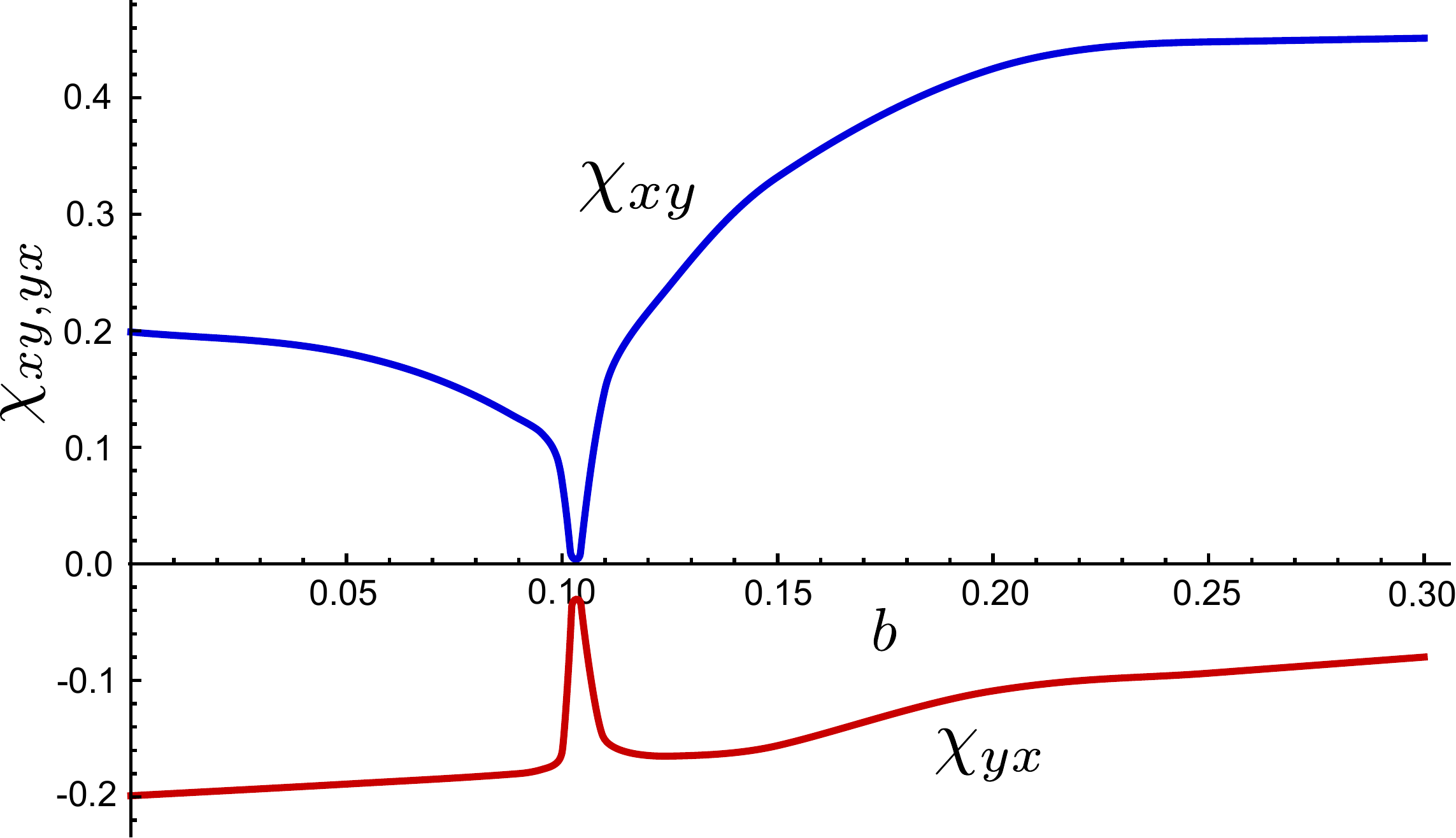}
\caption{(color online) The off-diagonal components of the Edelstein conductivity as functions of the magnetic field for the Fermi energy $\varepsilon_F=-0.8$.}
\label{fig6}
\end{figure}

As the Fermi energy increases, both components of the tensor $\chi_{ij}$ growth in the magnitude and then sharply drop when the Fermi level passes through the saddle point, as shown in Fig.~\ref{fig5}. Fig.~\ref{fig6} shows the dependence of $\chi_{xy}$ and $\chi_{yx}$ on the magnetic field.

The anisotropy of the Edelstein conductivity is manifested in the fact that the spin polarization vector is not perpendicular to the electric field that induces it, and the angle between the spin polarization and the electric field depends on the Fermi level.

\section{Conclusion}\label{S_conclude}
We have provided a method of theoretical study of anisotropic transport in 2D electron gas within the framework of the semiclassical Boltzmann theory. This method allows one to exactly find the nonequilibrium distribution function without using the relaxation time approximation in the case in which electrons are scattered by impurities with a short-ranged potential at zero temperature. In the presence of a complex structure of the Fermi contours, all possible electronic transitions are taken into account in calculations using this method, both within each contour and between different contours.

The method has been used to study anisotropic transport in a 2D electron gas with a SOI subjected to an in-plane magnetic field. In this case, the most important factor that determines the transport features is the Van Hove singularity of the density of states, which appears due to a combined effect of the magnetic field and the SOI\@. The singularity is controlled by the magnetic field. It appears at the bottom of the conduction band and rises to the Dirac point with increasing magnetic field. The Fermi contours are strongly anisotropic especially when the Fermi energy is close to the Van Hove singularity. The most interesting effects arise when the singularity passes through the Fermi level. 

Calculations of the conductivity tensor and the spin polarization induced by an in-plane electric field revealed two main effects that arise due to the interplay of the SOI and the magnetic field.

First, both the electrical conductivity and the Edelstein conductivity drop sharply when the Fermi level crosses the Van Hove singularity point. The presence of a minimum of electrical conductivity in 2D systems with a Van Hove singularity has been known for a long time~\cite{PhysRevB.53.11344}, and it continues to attract a great deal of interest~\cite{PhysRevLett.120.076602}. Our calculations are qualitatively consistent with the results known for other systems.

An interesting conclusion from our study of the conductivity minimum is that the magnetic field, at which the minimum is attained, is related to the parameters of the system studied here by a simple relation:
\begin{equation}
 B=\frac{m \alpha^2}{g\mu_B \hbar^2} +\frac{2\mu}{g\mu_B},
\end{equation}
where $\mu$ is the chemical potential controlled by external conditions. This fact can be used, for example, to determine the SOI constant from experiments.

Another interesting and unexpected result is that the anisotropy of the electric conductivity and the Edelstein conductivity is strongly changed as the Fermi level passes through the Van Hove singularity. In the energy region below the Dirac point, the conductivity anisotropy changes the sign twice when the Fermi level or magnetic field are changed. The anisotropy of the Edelstein conductivity leads not only to the angular dependence of the amplitude of the spin polarization, but also to the rotation of the spin polarization vector with respect to the electric field. Study of the transport anisotropy as a function of the Fermi energy or magnetic field in experiments can provide information on the anisotropy of the density of electronic states and scattering processes.

\acknowledgments
This work was partially supported by the Russian Foundation for Basic Research (Grant No.~17-02-00309) and the Russian Academy of Sciences. We are grateful to the referee, who drew our attention to the thesis~\cite{chesiPHD2007}.

\bibliography{anisotropy}

\end{document}